# SiZer for Censored Density and Hazard Estimation [*]

| Jiancheng Jiang | J. S. Marron |
|---|---|
| Department of Mathematics and Statistics | Department of Statistics |
| University of North Carolina | University of North Carolina |
| Charlotte, NC 28223, USA | Chapel Hill, NC 27599, USA |

June 06, 2008

**Abstract**

The SiZer method is extended to nonparametric hazard estimation and also to censored density and hazard estimation. The new method allows quick, visual statistical inference about the important issue of statistically significant increases and decreases in the smooth curve estimate. This extension has required the opening of a new avenue of research on the interface between statistical inference and scale space.

## 1 Introduction

Nonparametric hazard rate estimation is a standard tool in survival analysis, dating back at least to Watson and Leadbetter (1964a,b) and Rice and Rosenblatt (1976).

For practical use, a critical issue is understanding where the hazard rate curve increases and where it decreases. A confounding issue is the bandwidth, i.e. the window width or smoothing parameter. SiZer addresses both of these problems, in the context of nonparametric density and regression estimation, by combining a scale space approach to smoothing with a useful visualization of simultaneous statistical inference.

---

[*]J. S. Marron was supported by NSF Grant DMS-9971649.

The extension of SiZer developed in this paper is not straightforward. The obvious idea of simply plugging reweighted data into SiZer gives invalid statistical inference. Hence, a careful statistical accounting for the reweighting is developed. Some examples are shown in Section 2. Mathematical development is given in Section 3. Computational issues are discussed in Section 4.

It is straightforward to simultaneously develop these ideas for (i) hazard estimation; (ii) censored density estimation; (iii) censored hazard estimation. This is because all three of these cases fit very simply into a general form of estimator, using an elegant common notation, perhaps first published by Patil (1990, 1993). Hence all three cases are treated simultaneously here. For reasons of presentation, various aspects of this paper are usually illustrated by focusing on just one of the three cases first.

Some other important related references include Tanner and Wong (1983), Marron and Padgett (1987), Lo, Mack and Wang (1989), Sarda and Vieu (1991), Müller and Wang (1994), González-Manteiga, Marron, and Cao (1996), Kousassi and Singh (1997), Stute (1999), Hess, Serachitopol and Brown (1999), and Jiang and Doksum (2003).

Many readers perhaps understand that when data are censored, it is important to properly adjust for the bias that this creates, using Kaplan-Meier weights. An example, which shows that this bias can seriously impact SiZer inference, can be found in Section 1.1 of Jiang and Marron (2003). Kaplan-Meier weights, and there application in curve estimation, can be thought of in several ways. An insightful view is to consider the data in time order. All uncensored observations, that appear before any censored observations, receive equal weight. After the first censored observation appears, later uncensored observations receive more weight, etc. The intuitive content of the resulting Kaplan Meier weights is that the later uncensored observations, can be thought of as representing both themselves and also a fraction of the earlier censored observations in the data set. An example, which illustrates how letting one data point represent several can seriously im-pact SiZer inference, is given in Section 1.2 of Jiang and Marron (2003), in the somewhat different context of data rounding. See Figure 5 of Chaudhuri and Marron (1999) for another illustration of the data rounding phenomenon. If re-weighted data are simply plugged into SiZer, then treating the later uncensored observations as several data points, will affect the SiZer map in the same way, creating invalid inference. This problem is addressed by revising the SiZer inference, through a careful variance calculation, which results in correct inference.

## 2  Examples

Figure 1 shows a censored SiZer analysis of the Stanford Heart Transplant Data, from Kalbfleisch and Prentice (1980). The data consisting of 103 observations, originally from Crowley and Hu (1977) are the survival times (in days) of potential heart transplant recipients from their date of acceptance into the transplant program. There is censoring since some patients were lost to follow up before they died and since some patients were still alive on the closing date of the study.

Analysis from the point of view of density estimation is shown in Figures 1a and 1c. This shows that there are many deaths very soon after transplantation, and a long decreasing tail. Because of the relatively poor way in which the kernel density estimator handles boundaries, see e.g. Figure 2.16 of Wand and Jones (1995), at larger scales the estimates first increase at the left edge. SiZer shows that both the overall decrease (the large red region) is statistically significant, and so are the boundary effects (the thin blue region right at the edge).

For these data there is more interest in analysis of the hazard rate, as done in Figures 1b and 1d. The hazard rate is carefully defined in Section 3, but the intuitive idea is the instantaneous rate at which patients die. The estimate is a reweighting of the kernel density estimate, as can be seen from the fact that the small scale spikes in Figure 1b are simply magnifications of those in Figure 1a, but the scale is more appropriate for survival considerations. A central question is: when is the hazard rate increasing, and when is it decreasing? The color scheme of SiZer is well suited to address this issue. Furthermore, this question is much more directly answered by the SiZer approach, than by more conventional confidence intervals. The red in the middle left of the SiZer map in Figure 1d shows that the hazard rate significantly decreases during that time period, i.e. as transplants "settle in", chances of survival increase. The red near the top on the right shows that there is also a longer term improvement of the chances of survival after one has survived for a substantial period.

These findings are consistent with those of Jiang and Doksum (2003). An inconsistency is the blue region at the left end. As noted above this is due to poor boundary behavior of the kernel density estimator that underlies this inference. The local polynomial estimator developed by Jiang and Doksum (2003) avoids this problem, which is why their hazard estimate is mostly monotonically decreasing. An interesting open problem is to adapt SiZer ideas to the Jiang and Doksum local polynomial hazard estimator.

Figure 2 shows a SiZer analysis of the device lifetime data of Aarset(1987). These data are uncensored and consist of 102 observations.

The density estimates in Figure 2a suggest a "U-shape" density. However the SiZer map in Figure 2c flags only the right hand peak as statistically significant. This is likely due to the same inefficiency of the kernel density estimator near the boundary.

Of more interest for these data is the hazard rate analysis shown in Figures 2b and 2d. The dominant color in the SiZer map is blue which shows that the hazard rate generally increases over time, which is consistent with the expected wearing over time of mechanical components. A disappointing feature of the family of hazard estimates in Figure 2b, is that there is a spike only on the right side, while other analyses, including Aarset(1987) and Mudholkar, Srivastava and Kollia (1996) find a "bathtub" shape, that includes a spike on the left as well. This again is because of the poor boundary behavior of the kernel density estimator. This problem could also be addressed by a version of hazard estimation SiZer that is based on the local polynomial method of Jiang and Doksum

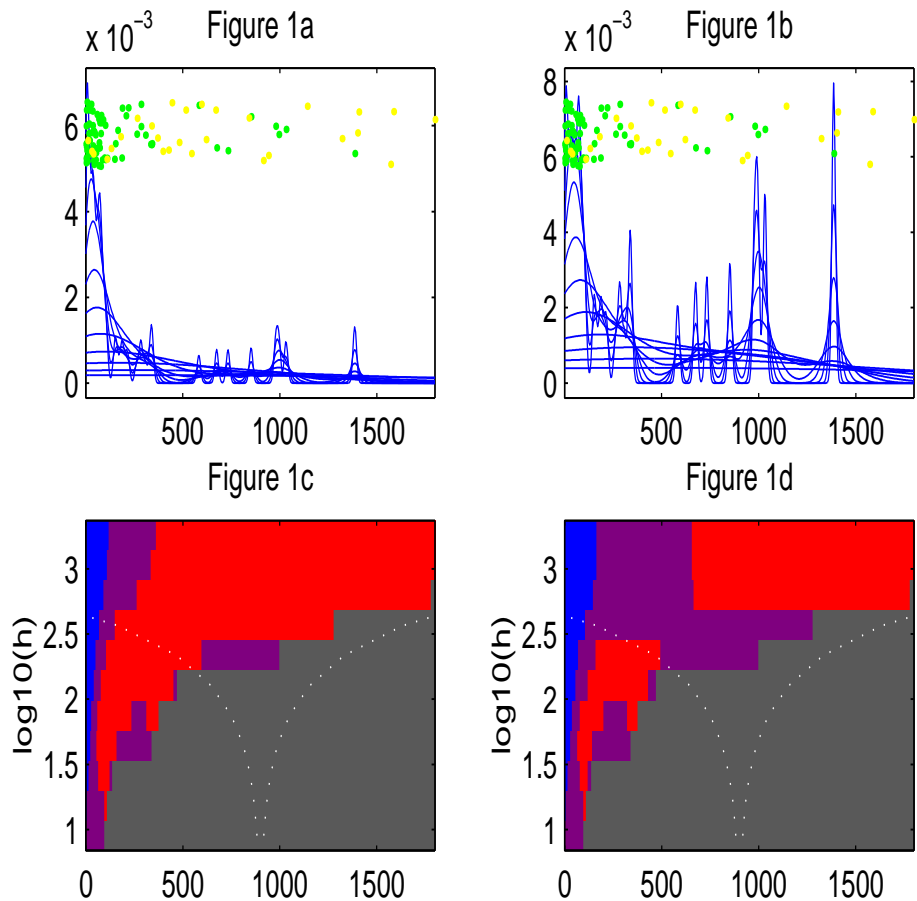

Figure 1: Days to death after heart transplant. Jitter plot, and family of censored density estimates in Figure 1a, with corresponding SiZer map in Figure 1c. Family of censored hazard estimates in Figure 1b, with corresponding SiZer map in Figure 1d.

(2003).

Both of these examples illustrate an important property of SiZer: it provides a generally good big picture assessment for initial exploratory purposes. However, for addressing any specific problem, e.g. the boundary questions brought up in Figures 1 and 2, it may not be as effective as a method that specifically targets that issue (although we do not know of a currently implemented method that gives better statistical inference of this type at the boundary). Hence we propose SiZer as a broad based method for initially finding structure in data (and for the perhaps more important task of quickly understanding what structures are mere sample artifacts). After one has an idea about what to look for, then other methods can provide deeper insights. Often the next useful step is modelling, e.g. as done by Mudholkar, Srivastava and Kollia (1996) for the device lifetime data.

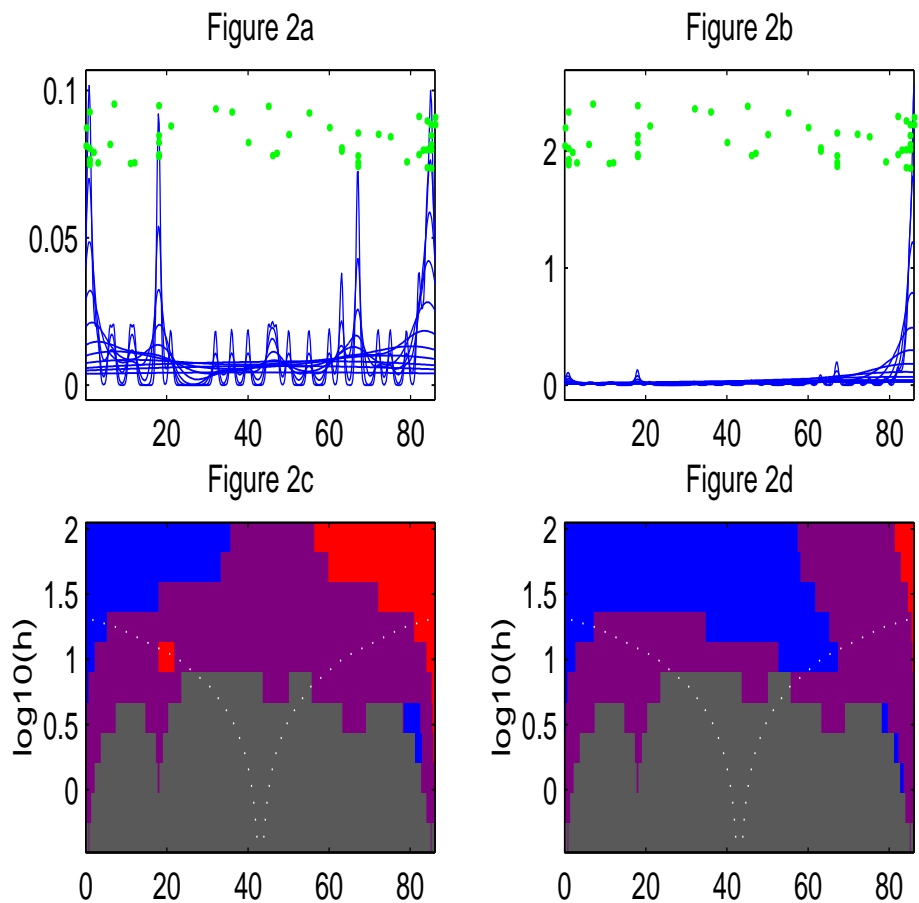

Figure 2: Device lifetime data. Family of censored density estimates in Figure 2a, with corresponding SiZer map in Figure 2c. Family of censored hazard estimates in Figure 2b, with corresponding SiZer map in Figure 2d.

## 3 Mathematical Development

Our extension of SiZer is most transparently explained in the context of hazard rate estimation. Hence this is developed in Section 3.1. Then the extension to censored density and censored hazard estimation is done in Section 3.2.

## 3.1 Hazard Rate Mathematics

For data $X_1, ..., X_n$ independent, identically distributed with cumulative distribution function $F(x)$, and probability density $f(x) = F'(x)$, the maximum likelihood estimate of $F$ is the empirical cumulative distribution function $F_n(x)$. The kernel density estimate of $f$ is

$$\widehat{f}_h(x) = n^{-1} \sum_{i=1}^{n} K_h(x - X_i),$$

where $K_h(\cdot) = \frac{1}{h} K\left(\frac{\cdot}{h}\right)$, for the kernel function $K$ and the bandwidth $h$. For $f$ supported on $(0, \infty)$ the hazard rate is $\lambda(x) = f(x)/[1 - F(x)]$, and its cumulative is $\Lambda(x) = \int_0^x \lambda(u)\, du$. Watson and Leadbetter (1964a) showed, but see for example Proposition 1 of Shorack and Wellner (1986) for a much different way to arrive at the same conclusion, that a natural estimate of the hazard rate is

$$\widehat{\lambda}_h(x) = n^{-1} \sum_{i=1}^{n} \frac{K_h(x - X_i)}{1 - F_n(X_i)} \qquad (1)$$

$$= \sum_{i=1}^{n} \frac{K_h\left(x - X_{(i)}\right)}{n - i},$$

where the $X_{(i)}$ are the order statistics, with $X_{(1)} \leq \cdots \leq X_{(n)}$.

Derivatives are estimated by differentiation. The density derivative, $f'(x)$, is estimated by

$$\widehat{f}'_h(x) = n^{-1} \sum_{i=1}^{n} K'_h(x - X_i), \qquad (2)$$

and the hazard rate derivative, $\lambda'(x)$, is estimated by

$$\widehat{\lambda}'_h(x) = n^{-1} \sum_{i=1}^{n} \frac{K'_h(x - X_i)}{1 - F_n(X_i)} = \sum_{i=1}^{n} \frac{K'_h\left(x - X_{(i)}\right)}{n - i},$$

where

$$K'_h(x) = \frac{\partial}{\partial x} K_h(x) = \frac{1}{h^2} K'\left(\frac{x}{h}\right).$$

The variance of the density derivative estimate is:

$$var\left(\widehat{f}'_h(x)\right) = var\left(n^{-1} \sum_{i=1}^{n} K'_h(x - X_i)\right) = n^{-1} var\left(K'_h(x - X_i)\right).$$

Denote by $s^2(y_1,\ldots,y_n) = n^{-1}\sum_{i=1}^n y_i^2 - (n^{-1}\sum_{i=1}^n y_i)^2$ and $K'_{0,i} = K'_h(x - X_i)$. Then the variance factor above is estimated by the sample variance

$$s^2\left(K'_{0,1}, ..., K'_{0,n}\right) = n^{-1}\sum_{i=1}^n \left(K'_{0,i}\right)^2 - \left(n^{-1}\sum_{i=1}^n K'_{0,i}\right)^2 \qquad (3)$$

$$= n^{-1}\sum_{i=1}^n \left(K'_{0,i}\right)^2 - \left(\widehat{f}'_h(x)\right)^2. \qquad (4)$$

Using the approximation $F_n(x) \approx F(x)$, the variance of the derivative hazard rate is approximated by

$$var\left(\widehat{\lambda}'_h(x)\right) \approx var\left(n^{-1}\sum_{i=1}^n \frac{K'_h(x - X_i)}{1 - F(X_i)}\right)$$

$$= n^{-1} var\left(\frac{K'_h(x - X_i)}{1 - F(X_i)}\right).$$

Except for the fact that $F$ is unknown the variance factor here could be estimated by the sample variance

$$s^2\left(K'_{F,1}, ..., K'_{F,n}\right),$$

where for any cumulative distribution function $H(x)$, dependence on $x$ and $h$ is suppressed in the notation

$$K'_{H,i} = \frac{K'_h(x - X_i)}{1 - H(X_i)}. \qquad (5)$$

Applying $F_n(x) \approx F(x)$, we get the approximation

$$s^2\left(K'_{F_n,1}, ..., K'_{F_n,n}\right). \qquad (6)$$

This is an important point where there is a critical difference between this development, and simply using the reweighted data in ordinary SiZer. In particular, the variance factor (6), now appropriately uses the weights. Thus, an isolated point with a heavy weight is no longer flagged as significant, because the variance estimate also increases when the weights are heavier.

SiZer gets its "simultaneous inference" properties (i.e. it addresses the multiple comparison problem) using a "number of independent blocks" calculation done in Section 3 of Chaudhuri and Marron (1999). The basis of this is the Effective Sample Size:

$$ESS_h(x) = \frac{\sum_{i=1}^{n} K_h(x - X_i)}{K_h(0)}, \qquad (7)$$

which measures the "number of points in each kernel window" (this is exactly true if $K$ is the uniform density window). Correct adaptation to the hazard context requires yet another careful twist. Naive reweighting would suggest that denominators of $1 - F(X_i)$ should be inserted. But the independent blocks calculation is based on the number of *independent* pieces of information, so instead the formula (7) should be retained in the same form.

Thus a hazard rate version of SiZer comes from modifying the density estimation version, replacing the terms $K'_h(x - X_i)$ in (2) by $K'_h(x - X_i)/[1 - F_n(X_i)]$, and replacing the variables $K'_{0,i}$ in (3) by $K'_{F_n,i}$.

## 3.2 Censored Estimation Mathematics

Censored data comes in the form $(X_1, \delta_1), ..., (X_n, \delta_n)$, where $X_i = \min(T_i, C_i)$ and $\delta_i = 1(T_i \leq C_i)$, where $1(A)$ is the indicator of event $A$ which equals 1 if $A$ happens and 0 otherwise.

Assume that $T_1, ..., T_n$ are independent, identically distributed with cumulative distribution function $F(x)$ and that $C_1, ..., C_n$ are independent (and independent of the $T_i$), identically distributed with cumulative distribution function $G(x)$. Note that the cumulative distribution function of $X_i$, is $L(x)$, where the corresponding cumulative survival function is $\overline{L}(x) = \overline{F}(x)\overline{G}(x)$, using the notation $\overline{H}(x) = 1 - H(x)$, for any cumulative distribution function $H(x)$.

The goal is estimation of the survival probability density $f(x) = F'(x)$ and the corresponding hazard rate $\lambda(x)$.

The cumulative distribution functions $F$ and $G$ can be estimated by the Kaplan Meier

(1958), i.e. Product Limit, estimators given by $\overline{F}_n$ and $\overline{G}_n$ respectively, where

$$\overline{F}_n = \begin{cases} 1 - \prod_{X_{(i)} \leq x} \left(\frac{n-i}{n-i+1}\right)^{\delta_{(i)}}, & \text{if } x \leq X_{(n)} \\ 0, & \text{if } x > X_{(n)} \end{cases},$$

and $\overline{G}_n$ is defined similarly to $\overline{F}_n$ but with $\delta_{(i)}$ replaced by $1 - \delta_{(i)}$, and the $(X_{(i)}, \delta_{(i)})$ are the order statistics version of the data with $X_{(1)} \leq \cdots \leq X_{(n)}$.

A natural kernel density estimate of $f$ is

$$\widehat{f}_h(x) = n^{-1} \sum_{i=1}^n \frac{\delta_i K_h(x - X_i)}{\overline{G}_n(X_i)}.$$

For $f$ supported on $(0, \infty)$ the corresponding estimate of the hazard rate is

$$\widehat{\lambda}_h(x) = n^{-1} \sum_{i=1}^n \frac{\delta_i K_h(x - X_i)}{\overline{G}_n(X_i) \overline{F}_n(X_i)}$$
$$= \sum_{i=1}^n \frac{\delta_{(i)} K_h(x - X_{(i)})}{n - i}.$$

Note that these have a structure very similar to the hazard rate estimator (1), which is why it is straight forward to extend SiZer to these cases as well.

Derivatives are again estimated by differentiation. The density derivative, $f'(x)$, is estimated by

$$\widehat{f}'_h(x) = n^{-1} \sum_{i=1}^n \frac{\delta_i K'_h(x - X_i)}{\overline{G}_n(X_i)}.$$

The hazard rate derivative, $\lambda'(x)$, is estimated by

$$\widehat{\lambda}'_h(x) = n^{-1} \sum_{i=1}^n \frac{\delta_i K'_h(x - X_i)}{\overline{L}_n(X_i)}.$$

The variance of the density derivative estimate is:

$$var\left(\widehat{f}'_h(x)\right) = var\left(n^{-1} \sum_{i=1}^n \frac{\delta_i K'_h(x - X_i)}{\overline{G}_n(X_i)}\right) = n^{-1} var\left(\frac{\delta_i K'_h(x - X_i)}{\overline{G}_n(X_i)}\right),$$

and for the hazard rate

$$var\left(\widehat{\lambda}'_h(x)\right) = var\left(n^{-1} \sum_{i=1}^n \frac{\delta_i K'_h(x - X_i)}{\overline{L}_n(X_i)}\right) = n^{-1} var\left(\frac{\delta_i K'_h(x - X_i)}{\overline{L}_n(X_i)}\right).$$

Using the approximation methods leading to (6), these variance factors are estimated by

$$s^2 \left( \delta_1 K'_{G_n,1}, ..., \delta_n K'_{G_n,n} \right), \tag{8}$$

using again the notation (5), and by

$$s^2 \left( \delta_1 K'_{L_n,1}, ..., \delta_n K'_{L_n,n} \right) \tag{9}$$

respectively.

The Effective Sample Size follows in a similar spirit. Again the basis is the number of independent pieces of uncensored data, resulting in the formula

$$ESS_h(x) = \frac{\sum_{i=1}^n \delta_i K_h(x - X_i)}{K_h(0)}.$$

Thus the censored density and censored hazard rate version of SiZer come from modifying the density estimation version, replacing the terms $K'_h(x - X_i)$ in (2) by $\delta_i K'_h(x - X_i) / \overline{G}_n(X_i)$ and $\delta_i K'_h(x - X_i) / \overline{L}_n(X_i)$ respectively, and by replacing the variables $K'_{0,i}$ in (3) by $K'_{G_n,i}$ and $K'_{L_n,i}$ respectively.

## 4  Fast Computation

Because SiZer relies on a large number of smooths, it is important to use a fast computational method. Several such are discussed by Fan and Marron (1994). The binned approach is especially well suited to SiZer.

Details of the binned implementation of $\widehat{f}'_h(x)$ are similar to those given in Chaudhuri and Marron (1999), which are based on those of Fan and Marron (1994), except that the kernels are now divided by appropriate cumulative distribution functions. In particular, for the equally spaced grid of points $\{x_j : j = 1, ..., g\}$, let the corresponding bin counts (computed by some method, we have always used the "linear binning" described in Fan and Marron (1994)) be $\{c_{0,j} : j = 1, ..., g\}$. Then for density SiZer

$$\widehat{f}'_h(x_j) \approx n^{-1} \overline{S}'_0(x_j),$$

where $\overline{S}'_0(x_j) = \sum_{j'=1}^{g} \kappa'_{j-j'} c_{0,j'}$ and $\kappa'_{j-j'} = K'_h(x_j - x_{j'})$. The approximated standard deviation of $\widehat{f}'_h(x_j)$, is

$$\widehat{sd}(x_j) = n^{-1/2} \sqrt{n^{-1} \sum_{j'=1}^{g} \left(\kappa'_{j-j'}\right)^2 c_{0,j'} - \left(n^{-1} \sum_{j'=1}^{g} \kappa'_{j-j'} c_{0,j'}\right)^2}.$$

The censored and hazard versions of SiZer require reconsideration of the linear binning algorithm. When a data point $X_i$ is between grid points $x_j$ and $x_{j+1}$, linear binning assigns weight

$$w_{i,j} = \frac{X_i - x_j}{x_{j+1} - x_j}$$

to the bin centered at $x_j$, and weight

$$w_{i,j+1} = \frac{x_{j+1} - X_i}{x_{j+1} - x_j}$$

to the bin centered at $x_{j+1}$, and weight 0 to all other bins. These result in bin counts

$$c_{0,j} = \sum_{i=1}^{n} w_{i,j}.$$

For a generic estimated cumulative distribution function $H_n$, these bin counts are replaced by

$$c_{H_n,j} = \sum_{i=1}^{n} \frac{w_{i,j} \delta_i}{\overline{H}_n(X_i)}.$$

This results in the binned approximation to the generic estimator:

$$n^{-1} \sum_{i=1}^{n} \frac{\delta_i K'_h(x - X_i)}{\overline{H}_n(X_i)} \approx n^{-1} \overline{S}'_{H_n}(x_j),$$

where

$$\overline{S}'_{H_n}(x_j) = \sum_{j'=1}^{g} \kappa'_{j-j'} c_{H_n,j'},$$

To similarly approximate $\widehat{sd}$, use

$$\widehat{sd}(x_j) = n^{-1/2} \sqrt{n^{-1} \sum_{j'=1}^{g} \left(\kappa'_{j-j'}\right)^2 c_{H_n,j'} \left(\frac{c_{H_n,j'}}{\widetilde{c}_{0,j'}}\right) - \left(n^{-1} \sum_{j'=1}^{g} \kappa'_{j-j'} c_{H_n,j'}\right)^2},$$

where the factor of $\left(\frac{c_{H_n,j'}}{c_{0,j'}}\right)$ in the second moment term gives the second factor of $\frac{1}{H_n}$ that appears in the second moment. Finally, the binned version of the Effective Sample Size needs to be based on the *unadjusted* bin counts of the uncensored data

$$\overline{ESS} = \frac{\sum_{j'=1}^{g} \kappa_{j-j'} \widetilde{c}_{0,j'}}{K_h(0)},$$

where $\widetilde{c}_{0,j} = \sum_{i=1}^{n} w_{i,j} \delta_i$ and $\kappa_{j-j'} = K_h(x_j - x_{j'})$.

## 5 Computational details

The results in this paper were obtained using Matlab. The codes are available from Steve Marron at

http://www.stat.unc.edu/postscript/papers/marron/software/

## 6 Discuss

We extend the SiZer to censored density and hazard rate estimation. This extension requires a statistical accounting for the reweighting. Our censored density estimation method is obviously applicable to un-censored cases. The SiZer of density and hazard rate estimate allows one to make quick, visual statistical inferences on the issue of statistically significant increases and decreases in the smooth density and hazard rate estimate.